\documentclass[twocolumn]{aastex61}
\usepackage{comment}

\usepackage[utf8]{inputenc}
\usepackage[T5,T1]{fontenc}

\submitjournal{ApJ}

\newcommand{\comments}[1]{}

\newcommand\kms{ km s$^{-1}$}

\newcommand\Msun{M$_\odot$}
\newcommand\myclump{G331.372-00.116}

\newcommand{\mum}{$\mu$m}

\shorttitle{IRDC G331.372-00.116}
\shortauthors{Contreras et al.}

\begin{document}


\title{Infall Signatures in a Prestellar Core embedded in the High-Mass 70 $\mu$m Dark IRDC G331.372-00.116}

\correspondingauthor{Yanett Contreras}
\email{ycontreras@strw.leidenuniv.nl}

\author{Yanett Contreras}
\affil{Leiden Observatory, Leiden University, PO Box 9513, NL-2300 RA Leiden, the Netherlands}

\author{Patricio Sanhueza}
\affil{National Astronomical Observatory of Japan, National Institutes of Natural Sciences, 2-21-1 Osawa, Mitaka, Tokyo 181-8588, Japan}

\author{James M. Jackson}
\affil{School of Mathematical and Physical Sciences, University of Newcastle, University Drive, Callaghan NSW 2308, Australia}

\author{Andr\'es E. Guzm\'an}
\affil{National Astronomical Observatory of Japan, National Institutes of Natural Sciences, 2-21-1 Osawa, Mitaka, Tokyo 181-8588, Japan}
\affil{Departamento de Astronom\'ia, Universidad de Chile, Camino el Observatorio 1515, Las Condes, Santiago, Chile}

\author{Steven Longmore}
\affil{Astrophysics Research Institute, Liverpool John Moores University, 146 Brownlow Hill, Liverpool L3 5RF, UK}

\author{Guido Garay}
\affil{Departamento de Astronom\'ia, Universidad de Chile, Camino el Observatorio 1515, Las Condes, Santiago, Chile}

\author{Qizhou Zhang}
\affil{Harvard-Smithsonian Center for Astrophysics, 60 Garden Street, Cambridge, MA 02138, USA}

\author{Quang Nguy{\fontencoding{T5}\selectfont \~\ecircumflex{}}n-Lu'o'ng}
\affil{The Canadian Institute for Theoretical Astrophysics (CITA), University of Toronto, 60 St. George Street, Toronto, Ontario, M5S 3H8, Canada}

\author{Ken'ichi Tatematsu}
\affil{National Astronomical Observatory of Japan, National Institutes of Natural Sciences, 2-21-1 Osawa, Mitaka, Tokyo 181-8588, Japan}

\author{Fumitaka Nakamura} 
\affil{National Astronomical Observatory of Japan, National Institutes of Natural Sciences, 2-21-1 Osawa, Mitaka, Tokyo 181-8588, Japan}

\author{Takeshi Sakai} 
\affil{The University of Electro-Communications, Chofu, Tokyo 182-8585, Japan}

\author{Satoshi Ohashi}
\affil{RIKEN, 2-1, Hirosawa, Wako-shi, Saitama 351-0198, Japan}

\author{Tie Liu}
\affil{Korea Astronomy and Space Science Institute, 776 Daedeokdaero, Yuseong-gu, Daejeon 34055, Republic of Korea}
\affil{East Asian Observatory, 660 N. A'ohoku Place, Hilo, HI 96720, USA}

\author{Masao Saito}
\affil{National Astronomical Observatory of Japan, National Institutes of Natural Sciences, 2-21-1 Osawa, Mitaka, Tokyo 181-8588, Japan}

\author{Laura Gomez}
\affil{Joint ALMA Observatory, Alonso de C\'ordova 3107, Vitacura, Santiago, Chile}

\author{Jill Rathborne}
\affil{CSIRO Astronomy and Space Science, P.O. Box 76, Epping NSW 1710, Australia}

\author{Scott Whitaker}
\affil{Physics Department, Boston University, Boston, MA, USA}

\begin{abstract}
Using Galactic Plane surveys, we have selected a massive (1200 M$_\odot$), cold (14 K) 3.6-70 $\mu$m dark IRDC G331.372-00.116. This IRDC has the potential to form high-mass stars and, given the absence of current star formation signatures, it seems to represent the earliest stages of high-mass star formation. We have mapped the whole IRDC with the Atacama Large Millimeter/submillimeter Array (ALMA) at 1.1 and 1.3 mm in dust continuum and line emission. The dust continuum reveals 22 cores distributed across the IRDC. In this work, we analyze the physical properties of the most massive core, ALMA1, which has no molecular outflows detected in the CO (2-1), SiO (5-4), and H$_2$CO (3-2) lines. This core is relatively massive ($M$ = 17.6 M$_\odot$), subvirialized (virial parameter $\alpha_{vir}=M_{vir}/M=0.14$), and is barely affected by turbulence (transonic Mach number of 1.2). Using the HCO$^+$ (3-2) line, we find the first detection of infall signatures in a relatively massive, prestellar core (ALMA1) with the potential to form a high-mass star. We estimate an infall speed of 1.54 km s$^{-1}$ and a high accretion rate of 1.96 $\times$ 10$^{-3}$ M$_\odot$ yr$^{-1}$. ALMA1 is rapidly collapsing, out of virial equilibrium, more consistent with competitive accretion scenarios rather than the turbulent core accretion model. On the other hand, ALMA1 has a mass $\sim$6 times larger than the clumps Jeans mass, being in an intermediate mass regime ($M_{J}=2.7<M\lesssim$ 30 M$_\odot$), contrary to what both the competitive accretion and turbulent core accretion theories predict.

\end{abstract}


\section{Introduction}
\label{intro}

Stars form from dense ($\sim$$10^6$ cm$^{-3}$), compact ($\sim$$0.05-0.1$ pc) self-gravitating regions, or `cores', that are embedded in extended, usually filamentary, clumps of molecular gas ($\sim$1 pc). While the formation of low-mass stars is relatively well understood, the formation of high-mass stars ($>$8 M$_\odot$) remains shrouded in uncertainties. One of the main questions about the formation of high-mass stars is how they gather their initial mass. To answer this, one requires to measure the initial condition of the gas within cores that have not yet formed high-mass stars. At later evolutionary stages, the outflows, stellar heating, and ionization from the newly formed stars will affect its environment making it more difficult to study their mass accretion mechanisms. 

Currently, two main scenarios aim to explain how high-mass stars acquire their mass (see review by \citet{Tan-2014}). In the `competitive accretion' theory \citep{bonnell-2001}, at early times a molecular cloud fragments into a swarm of small cores, each with a mass roughly equal to the thermal Jeans mass ($\sim$2 M$_\odot$ at a volume density of 5$\times$10$^4$ cm$^{-3}$ and a temperature of 12 K). As gas funnels down the gravitational potential well of the much larger molecular clump, those cores near the center of the potential receive a fresh supply of gas from afar and grow via Bondi-Hoyle accretion. In this scenario it is expected to see signatures of global collapse in the ambient gas, which would suggest that material is being funneling into the cores \citep[e.g.][]{Peretto-2013, Liu-2013, Liu-2016}. 

On the other hand, in the `turbulent core accretion' scenario \citep{mckee-tan-2003}, cores are fed only locally. In this theory, all of the mass accreted onto the final high-mass star originates locally, at birth, within the core \citep{Tan-2014}. One of the main observational predictions of `turbulent core accretion' is the presence of high-mass prestellar cores ($>$30 M$_\odot$, necessary to form a 8 M$_\odot$ star assuming a star formation efficiency of 30\%). However, there is a of lack of evidence regarding the existence of these prestellar cores. The best candidates found so far correspond to G11.920.61-MM2 \citep{Cyganowski-2014}, G11P6-SMA1 \citep{Wang-2014}, and C1-S \citep{Kong-2017}. G11.920.61-MM2 has a mass of $\sim$30 M$_\odot$ but lack emission from molecular lines, making it very peculiar. G11P6-SMA1 has a mass of 27.9 M$_\odot$ and shows no outflow emission making it a good pre-stellar core candidate. C1-S has a mass ranging between 11 and 53 M$_\odot$, due to the uncertainty in their temperature values, which range between 7 and 13 K.

In this paper, we present observations of the massive, 70 $\mu$m dark IRDC G331.372-00.116 obtained with the Atacama Large Millimeter/submillimeter Array (ALMA). This IRDC has all the characteristics of a high-mass stellar cluster candidate in the  prestellar phase. We present the ALMA 1.1 mm dust continuum emission for the whole cloud, but focus on the remarkable physical 
properties of the brightest core. Given the unprecedented angular resolution and sensitivity provided by ALMA at relatively high excitation 
transitions, we have detected infall signatures traced by HCO$^+$ (3-2) in a subvirialized prestellar, relatively massive core. 

\section{IRDC G331.372-00.116}
\label{irdc-prop}

\begin{figure*}
\begin{center}
\includegraphics[trim={1cm 0.5cm 5cm 6cm}, clip, angle=0,scale=0.57]{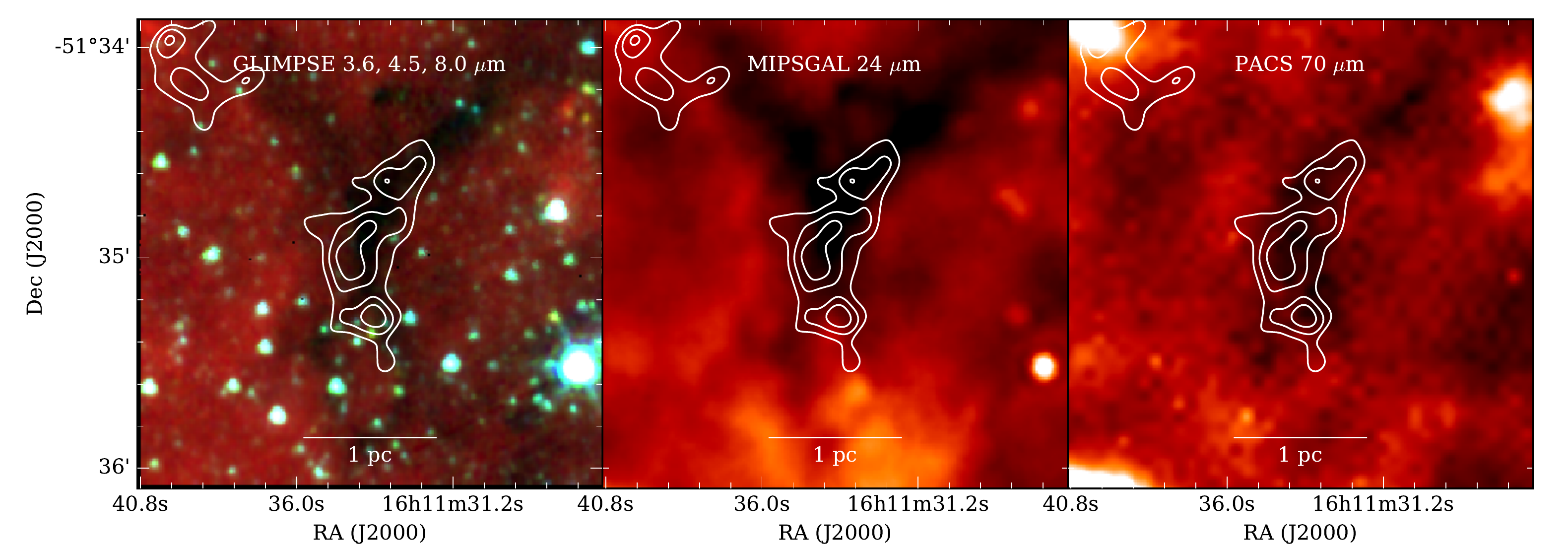}
\end{center}
\caption{Infrared emission toward \myclump. \myclump~appears as a dark feature even up to 70 $\mu$m, confirming the early stage of evolution of this clump. \textit{Left panel}: Three color image of the \textit{Spitzer} GLIMPSE bands (Blue: 3.6 $\mu$m, green: 4.5 $\mu$m, and red: 8 $\mu$m). \textit{Middle panel}: \textit{Spitzer} MIPSGAL 24 $\mu$m infrared emission. \textit{Right panel}: \textit{Herschel} PACS 70 $\mu$m emission. In all the panels the 70\% to 90\% of the peak dust continuum emission from ATLASGAL is overlaid in white contours. }
\label{g331_ir}
\end{figure*}

G331.372-00.116 is located in the Galactic plane, in the RCW 106 complex \citep{Nguyen-2015}, at a distance of 5.4 kpc \citep{Whitaker-2017}. It was detected via its bright continuum emission at 870 \mum~in the ATLASGAL survey \citep[AGAL331.372-00.116,][]{Contreras-2013a}. It appears as a dark silhouette against the bright diffuse IR background in the \textit{Spitzer} 3.6 $\mu$m, 4.5 \mum, 8 \mum\ GLIMPSE \citep{Churchwell-2009}, and 24 \mum\ MIPSGAL \citep{carey-2005} images (see Figure \ref{g331_ir}). It also appears as dark in the 70 \mum\ PACS Herschel image, suggesting that the temperature of this clump is low. Indeed, using dust continuum emission from Herschel PACS and SPIRE images, and ATLASGAL at 870 \mum,  \citet{Guzman-2015} derive a dust temperature for G331.372-00.116 of 14$\pm4$ K.

\myclump~consists of a single clump with a mass ($M$) of $1200$ M$_\odot$\footnote{This value differs from the 1640 M$_\odot$ derived in \citet{Contreras-2017} due to contamination in the dust emission from a second velocity component, undetected in the MALT90 data, located in the clump's upper region. Based on the C$^{18}$O emission, we estimate that the clump has a mass corresponding to the 75\% of the 1640 M$_\odot$ previously estimated.} and an effective radius of 0.56 pc, resulting in an average volume density of $2.4\times10^4$ cm$^{-3}$ and a surface density of 0.26 gr cm$^{-2}$ \citep{Contreras-2017}. Assuming a 30\% star formation efficiency, this clump should form a stellar cluster of 360 M$_\odot$. Making two different estimations following the empirical relation from \cite{Larson-2003} and the IMF from \cite{Kroupa-2001}, a stellar cluster of 360 M$_\odot$ should host a high-mass star of 17 and 24 M$_\odot$, respectively \citep[Equations 1 and 2 in ][]{Sanhueza-2017}. The virial mass ($M_{vir}$) of this clump is 760 M$_\odot$, based on the single-dish Mopra observations of the N$_2$H$^+$(1-0) transition with a line width of 3.0$\pm$0.2 km s$^{-1}$  \citep[value typically measured in IRDCs;][]{Sanhueza-2012} obtained from the Millimetre Astronomy Legacy Team 90 GHz (MALT90) Survey \citep{Rathborne-2016,Jackson-2013,Foster-2013,Foster-2011}. The virial parameter is $\alpha_{vir}=M_{vir}/M=0.6$, suggesting that the clump embedded in this IRDC is gravitationally bound. 

Overall, the massive IRDC G331.372-00.116 has all physical properties needed to form a stellar cluster containing at least one high-mass star. 

\section{Observations}
\label{alma-obs}

The observations were made with ALMA, located in the Llano de Chajnantor, Chile. We used two ALMA datasets obtained during ALMA Cycle 2 (Project ID: 2013.1.00234.S; PI: Gomez) and Cycle 3 (Project ID: 2015.1.01539.S; PI: Sanhueza) that covered the continuum and molecular line emission. 

The first dataset was obtained during June 2014 and May 2015. The 12 m array consisted of 34 antennas, with baselines ranging from 15 to 348 m. Large-scale continuum and line emission was recovered by including the Atacama Compact Array (ACA) using the 7 m and Total Power (TP) arrays. The 7 m array observations consisted of 9 antennas, with baselines ranging from 8 to 48 m. For both 12 and 7 m arrays, the flux calibration and phase referencing were carried out using J1427-421 and J1617-5848, respectively. For the TP observations flux calibration was done using Uranus. 

The whole IRDC was covered by a 12-pointing mosaic using each array (12 and 7 m). The angular resolution of the combined images is 1\farcs2 (0.03 pc at 5.4 kpc) and the largest angular scale recovered for the continuum emission corresponds to 39" (1 pc at 5.4 kpc). For the line emission, interferometric and single-dish (TP) data were combined.

The receiver setup corresponds to the Band 6 of ALMA centered at $\sim$267 GHz. We observed 16 spectral windows in dual polarization mode, with a bandwidth of 234 MHz. The velocity resolution of the spectral windows range between 0.5 and 1.2 km s$^{-1}$. 
Our setup targeted the optically thick transitions HCO$^+$ (3-2) and HNC (3-2) and covered the following molecular lines that were not detected: HC$^{18}$O$^+$ (3-2), the deuterated molecule NHD$_2$ 4(2,2)-4(1,3), and the high-excitation energy or shock-tracing lines H$_2$CS 8(1, 8)-7(1, 7), SO$_2$ 4(3-1)-4(2-2) and SO$_2$ 5(3,3)-5(2-4).

All data reduction was performed using the CASA software package \citep{McMullin-2007}. Each individual dataset was independently calibrated before being merged. The 12 and 7 m array datasets were concatenated and cleaned together using the CASA \texttt{tclean} algorithm with a Briggs weighting of 0.5. To avoid artifacts due to the complex structure seen in the continuum, we used a multi-scale clean, with scale values of 0, 3, 10 and 30 times the image pixel size (0.16"). To create the continuum image, all the line free channels were used. The 12 and 7 m array line emission was combined with the TP observations through the feathering technique. The achieved rms of the continuum is 0.06 mJy. For the lines we used the \texttt{yclean} script that automatically cleaned each map channel with custom made masks (see Appendix \ref{appendixyclean}). The achieved rms for the lines is 0.06 K at a velocity resolution of 0.5 km s$^{-1}$. 

The second dataset was obtained during January 2016 and June 2016. The 12  m array consisted of 48 antennas, with baselines ranging from 15 to 331 m. Large-scale continuum and line emission was also recovered thanks to the inclusion of the ACA and TP arrays. The 7 m array observations consisted of 8 antennas, with baselines ranging from 8 to 44 m. Flux calibration was done using Ganymede for the 12 m array observations, and Ganymede, J1256-0547 and J1924-2914 for the 7 m array observations. Phase referencing was done using J1603-4904 for the 12 and 7 m arrays. For the TP observations, flux calibration was done using Uranus. 

The IRDC was covered by a 10-point mosaic using the 12 m array and a 3-pointing mosaic using the ACA. The angular resolution of the images is 1\farcs2 and the largest angular scale recovered corresponds to 43" (1.1 pc at 5.4 kpc, for the continuum emission). 

The receiver setup corresponded to the Band 6 of ALMA centered at $\sim$224 GHz in dual polarization mode. 
The velocity resolution of the spectral windows ranged between 0.17 km s$^{-1}$ and 1.3 km s$^{-1}$.This setup covered the H$_2$CO (3-2), SiO (5-4), CO (2-1), N$_2$D$^+$ (3-2), DCO$^+$ (3-2), and C$^{18}$O (2-1) molecular lines. In this paper we focus in the N$_2$D$^+$ (3-2), DCO$^+$ (3-2), and C$^{18}$O (2-1) transitions.

All data reduction was performed following the same procedure as for the previous dataset. The achieved rms of the continuum is 0.1 mJy. For the lines the achieved rms is 0.14 K at a velocity resolution of 0.08 km s$^{-1}$. 

In what follows, all sensitivities given for both ALMA data sets and the analysis correspond to the combined array data sets (without the TP for the continuum emission). For the dust analysis we used only the Cycle 2 data at 267 GHz as it has better sensitivity.

\begin{figure*}[t]
\begin{center}
\includegraphics[trim={0cm 3cm 1.5cm 3cm}, clip,angle=-90,scale=0.73]{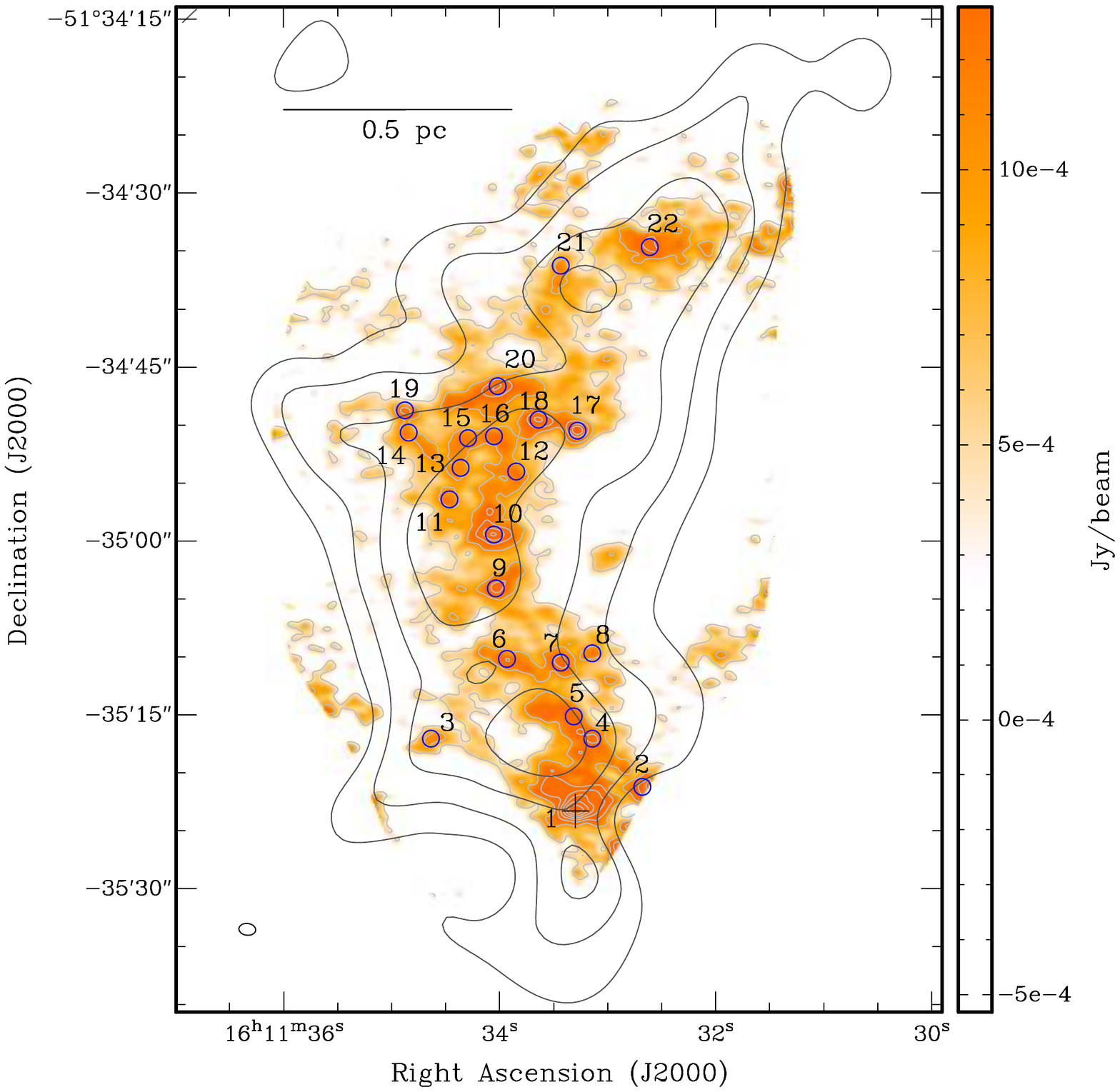}
\end{center}
\caption{Dust continuum emission at 1.1 mm (267 GHz) toward \myclump. Overlaid in contours is the ATLASGAL dust continuum emission (60\% to 90\% of the peak emission). The ALMA beam is shown in the lower left corner of the images. The cross shows the position of ALMA1 and the rest of the cores identified are marked by blue circles with their respective number. This images shows that the overall emission recovered by ALMA follows well the shape and elongation of the ATLASGAL emission. However, although there are several cores at the center of the map, the most massive core is located in the lower part of the IRDC, not being coincident with the peaks observed in the ATLASGAL images. We recovered the same core distribution on our ALMA observations carried out during 2013 and 2015. These observations were conducted independently, confirming that there is no error on the coordinates of the ALMA images.}
\label{g331_dust}
\end{figure*}

\section{Results}

\subsection{Dust Structure and Core Detection using Continuum Emission}

Figure \ref{g331_dust} shows the distribution of the 1.1 mm (267 GHz) dust continuum emission obtained with ALMA. 
A total of 22 cores are detected above the 5$\sigma$ level embedded in the $\sim$1.6 pc long filament. Cores are defined and their fluxes extracted by using the dendrogram technique \citep{Rosolowsky-2008}\footnote{To compute the dendrogram we used a threshold of 10 $\times$ rms, a step of 1 $\times$ rms, and a minimum number of pixels equals to the beam's area divided by the pixel size.}. Most of the cores are not isolated. They rather appear connected by more diffuse material that is recovered by the ACA array. The continuum 
emission revealed by ALMA generally follows the structure of the single-dish observations (elongation  and 
curvature). The central region of the IRDC fragments in several cores, but remarkably the brightest core in the 
whole IRDC is located in the southern region. The discrepancy seen between the single dish and ALMA observations is seen in both epochs of ALMA observations. It can be explained by the great amount of extended diffuse emission in this IRDC that was not recovered by ALMA. Indeed only 10\% of the flux observed by the single dish was detected by the 12 and 7 m ALMA arrays combined. 

The mass of the cores is calculated by using the following expression:
\begin{equation}
 M = \mathbb{R}~\frac{F_\nu D^2}{\kappa_\nu B_\nu (T)}~,
\label{eqn-dust-mass}
\end{equation}
where $F_\nu$ is the measured integrated source flux, $\mathbb{R}$ is the
 gas-to-dust mass ratio, $\rm D$ is the distance to the source, $\kappa_\nu$ is the dust opacity per gram of dust, and $B_\nu$ is the Planck function at the dust temperature $T$. Assuming a dust emissivity index ($\beta$) of 1.7 and scaling the value of 0.9 cm$^2$ g$^{-1}$  for $\kappa_{1.3 mm}$, which corresponds to the opacity of dust grains with thin ice
 mantles at gas densities of 10$^6$ cm$^{-3}$ \citep{Ossenkopf94}, we obtain $\kappa_{1.1 mm}$ = 1.13.   
A gas-to-dust mass ratio of 100 is assumed in this work. For the cores, we have adopted the {\it Herschel} dust temperature ($T$ = 14 K) determined at the clump scale by \cite{Guzman-2015}. Given that the temperature has been determined on scales much larger than the cores, this temperature represents a lower limit if deeply embedded star formation activity exists in the cores. If the cores are genuinely prestellar then the core temperature might be lower than this, and therefore the {\it Herschel} value would represent an upper limit. Using this dust temperature, we derive core masses ranging from 0.8 to 17.6 \Msun. Based on the uncertainty analysis of \cite{Sanhueza-2017}, 
we estimate that core masses, densities, and surface densities have $\sim$50\% uncertainty.
 
The number density is calculated by assuming a spherical core and using the molecular mass per hydrogen molecule ($\mu_{\rm H_2}$) of 2.8 \citep{Kauffmann08}. Table \ref{tbl-clumps} summarizes the physical properties derived for all the cores identified in \myclump. 

In this work, we focus on the physical properties of the most 
massive core, ALMA1, and defer the analysis of the whole IRDC to a future paper. 

\begin{figure}
\begin{center}
\includegraphics[trim={1.5cm 0cm 4cm 0cm}, clip, angle=0,scale=0.42]{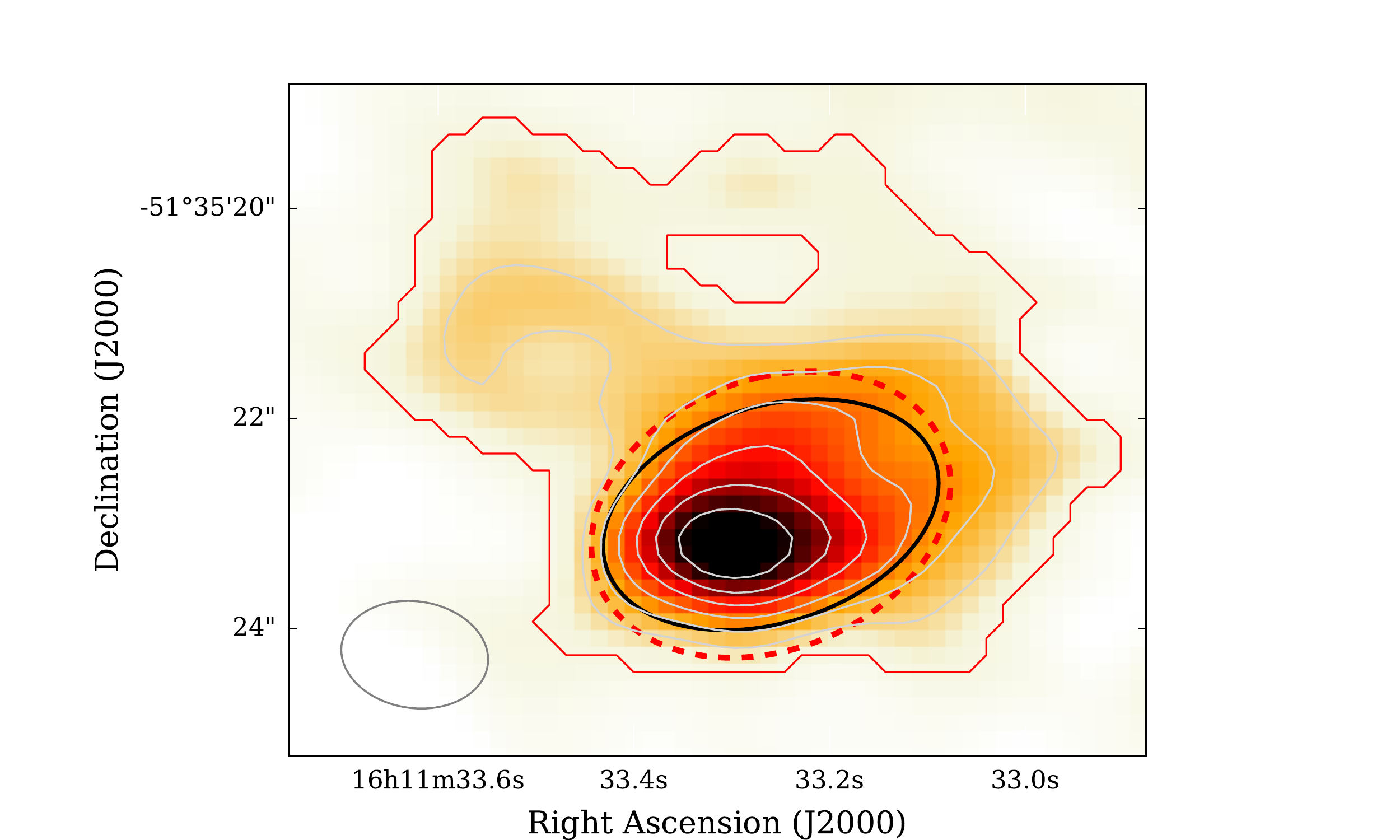}
\end{center}
\caption{Dust continuum emission at 1.1 mm toward ALMA1. Overlaid in contours is shown the 40\% to 90\% of the dust continuum emission. The red contour shows the area of the leaf defined by the dendrogram. The dashed red ellipse shows the area representing the first and second moments of the structure determined by the dendrogram. The black ellipse shows the 2-D Gaussian fit obtained with CASA. The beam is shown in the lower left corner of the image. }
\label{zoomalma1}
\end{figure}

Figure \ref{zoomalma1} shows the region surrounding ALMA1. The core is resolved at 1\farcs2 resolution. 
Using the dendrogram technique, we derive a flux and deconvolved 
FWHM size of 26.0 mJy and 3\farcs77$\times$2\farcs64, respectively. The geometric mean of the major and minor axis is 3\farcs15. The effective radius ($r_{eff}=\sqrt{A/\pi}$) 
 is 1\farcs57, i.e., 0.041 pc ($\sim$8500 AU) at the distance of 5.4 kpc, where $A$ is the area of the ellipse determined via the dendrogram. The mass ($M$), number density (n(H$_2$)), and surface density ($\Sigma$ = $M/(\pi r^2$)) of ALMA1 are 17.6 \Msun, 0.85$\times$10$^{6}$ cm$^{-3}$, and 0.65 gr cm$^{-2}$, respectively. At these scales, we would expect from ALMA1 to form a single or a few stars at most, as suggested from the fragmentation scales of $\sim2000$ AU seeing in numerical simulations \citep[e.g.,][]{Krumholz-2012}.
 
We compare the core properties derived by using dendrograms with the properties obtained by fitting a \mbox{2-D} Gaussian with the CASA software package. We obtain a  flux of 21.9$\pm1.6$ mJy and a deconvolved FWHM size of (3\farcs5$\pm$0\farcs3)$\times$(2\farcs1$\pm$0\farcs2), both comparable to the values obtained with dendrograms ($\sim$15\% different). Using the 2-D Gaussian values, we obtain a mass of 14.9 \Msun, 
density of 1.2 $\times$10$^{6}$ cm$^{-3}$, and surface density of 0.75 gr cm$^{-2}$. For this work, we have adopted the values obtained via the dendrogram technique. 

\subsection{Line Emission: Infall Signatures}

\begin{figure*}[t]
\begin{center}
\includegraphics[angle=0,scale=0.32, trim={0.1cm 0 3.0cm 0}, clip]{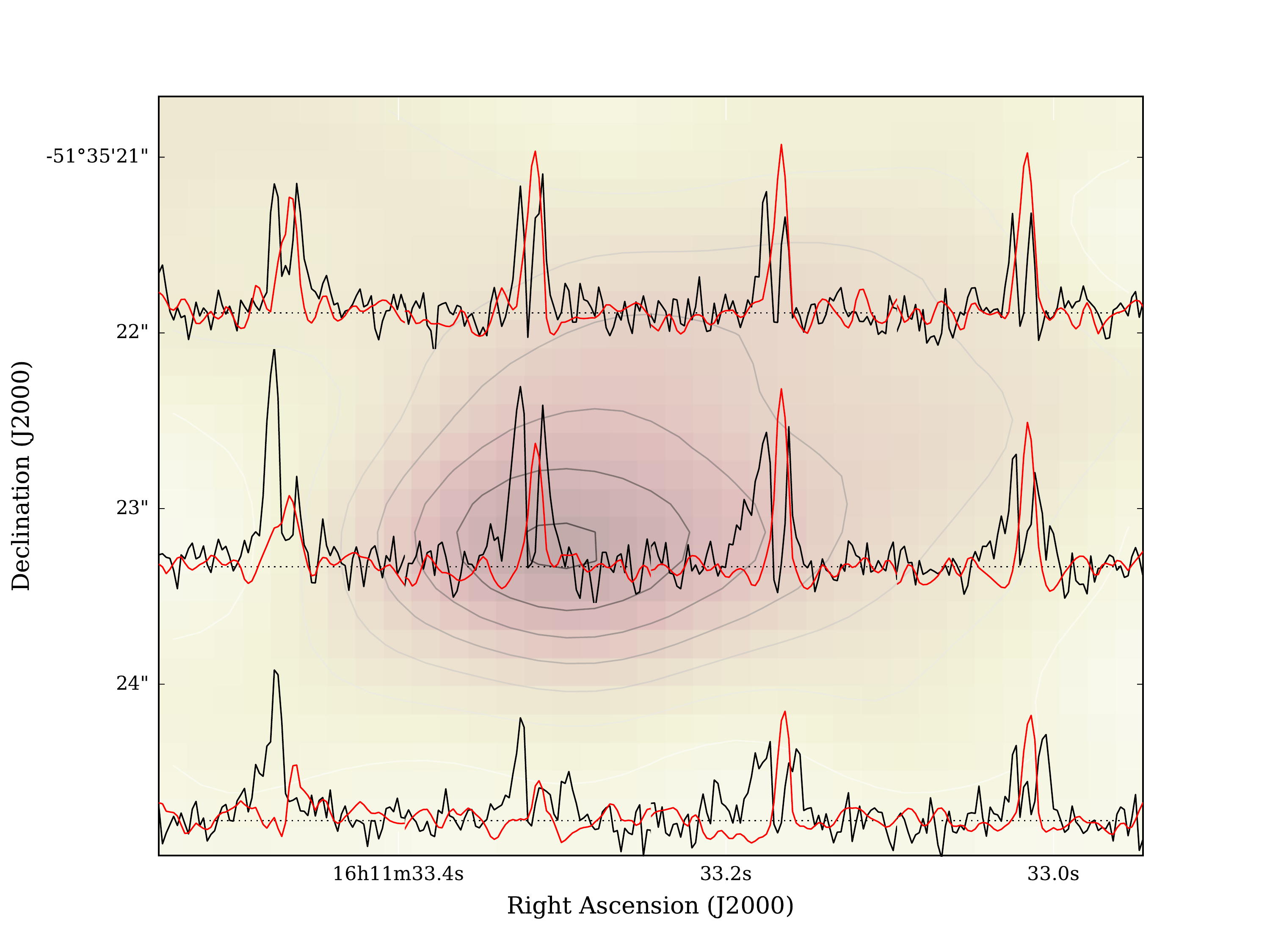}
\includegraphics[angle=0,scale=0.47, trim={0cm 0 2cm 0}, clip]{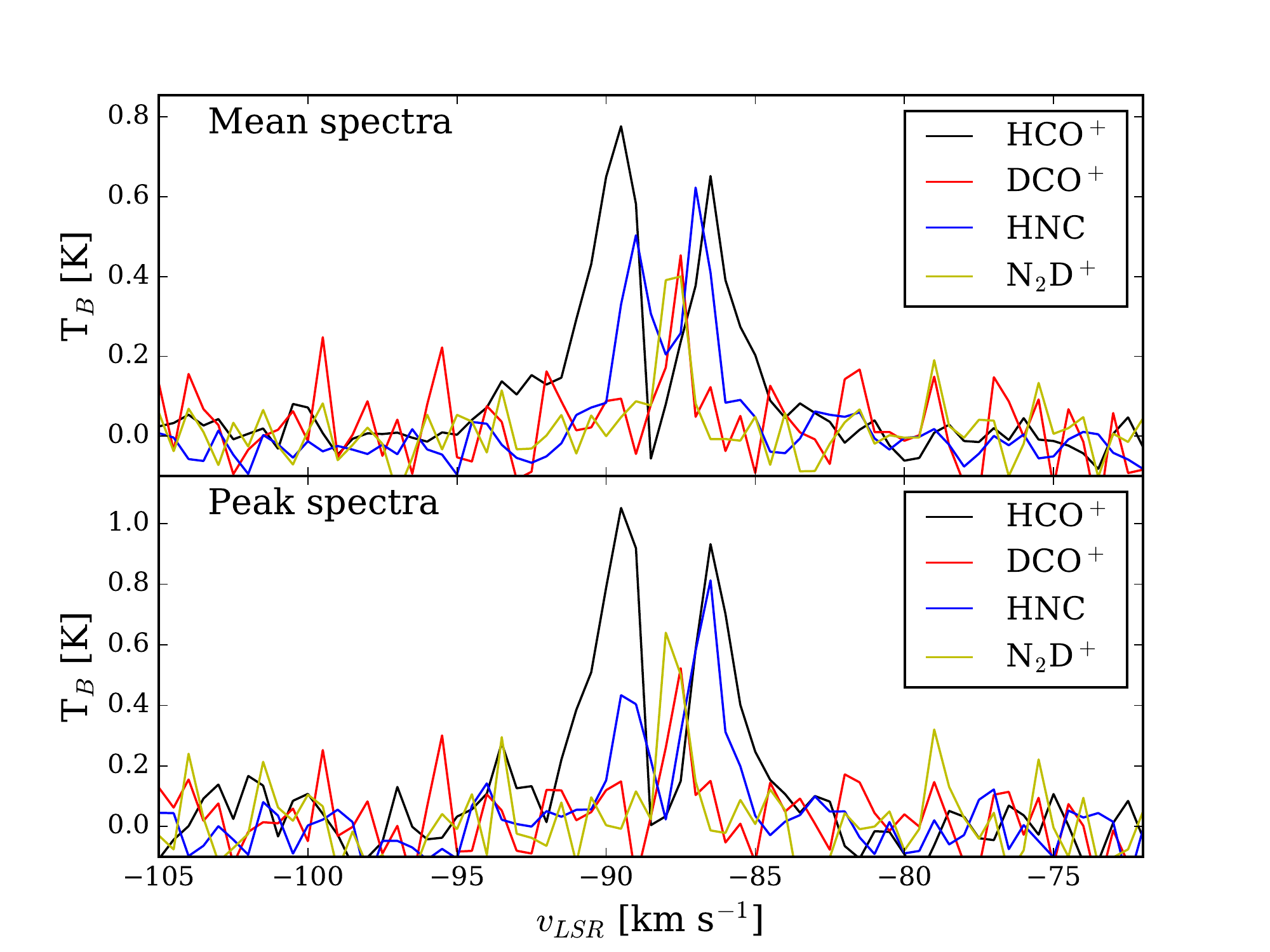}
\end{center}
\caption{Left: HCO$^+$ (3-2) (black) and C$^{18}$O (2-1) (red) spectra overlaid to the dust continuum emission observed towards ALMA1. Right upper panel: Mean spectra across ALMA1 of the HCO$^{+}$ (3-2), DCO$^+$ (3-2), HNC (3-2) and N$_2$D$^+$ (3-2) line emission. Here both optically thin tracers shows a single peak at the position of the $v_{LSR}$ of the core. The HNC mean spectra shows a small red-peaked profile, while HCO$^+$ shows clearly the typical blue-peaked infall profile. Right lower panel: Peak spectra of the same four lines as in the upper panel. Here the infall profile is also  evident toward the peak position, while for HNC there is no evidence of infall.}
\label{spec1}
\end{figure*}

Figure \ref{spec1} displays the observed HCO$^+$ (3-2) and C$^{18}$O (2-1) spectra across ALMA1, and the mean and peak HCO$^+$ (3-2), DCO$^+$ (3-2), HNC (3-2), N$_2$D$^+$ (3-2) spectra toward ALMA1. The 
HCO$^+$ and HNC profiles show a self-absorbed profile at the systemic v$_{lsr}$ (-87.7 \kms) of the core, which is determined by the central velocity of 
the deuterated molecules (N$_2$D$^+$ and DCO$^+$). The HCO$^+$ spectra show a blue-red asymmetry with the blue-shifted peak brighter than 
the red-shifted peak. This profile is typically explained as a sign of infall, assuming the inner gas in the core has a warmer excitation temperature  
than the envelope. The profile of the optically thin tracers (N$_2$D$^+$, DCO$^+$) peak at the velocity of the self-absorption,  confirming that the double-peaked 
profile is not due to two velocity components along the line of sight. A warmer excitation temperature in the inner region of the prestellar core can be explained by the 
high critical densities of the HCO$^+$ and HNC (3-2) transitions, 1.1$\times$10$^6$ and 2$\times$10$^6$ cm$^{-3}$, respectively. The core (average) density is 
 0.85$\times$10$^6$ cm$^{-3}$, similar to the critical density of the infall tracers. At the center of the core, the density can be higher and the HCO$^+$ and HNC emission can be thermalised; on the other hand, in the envelope the gas density can be lower than 10$^6$ cm$^{-3}$, making the emission from the J=3-2 transitions sub-thermal with lower excitation temperatures than at the center.  
Although the HNC line displays a self-absorbed profile similar to the HCO$^+$ line, the blue- and red-shifted peaks have similar intensity and sometimes the blue-red asymmetry is reversed with the red peak slightly brighter than the blue peak. Some works \cite[e.g.,][]{Redman-2004} suggest that core rotation and 
 outflow activity can produce both kinds of asymmetries. We rule out the presence of molecular outflows by inspecting the CO (2-1), SiO (5-4), and H$_2$CO (3-2) lines, as well as the HCO$^+$ and HNC (3-2) lines. We detect no high-velocity 
 gas that would indicate molecular outflows originated from ALMA1, and thus confirm its prestellar nature.  
 
 As seen in \cite{Redman-2004}, rotation would produce blue-red asymmetries with brighter 
 blue peaks on one side of the core's axis of rotation and brighter red peaks on the other side. This asymmetry pattern is not observed in ALMA1. Furthermore, no velocity gradients are detected in the optically thin tracers that could reveal signs of rotation. We suggest that the more likely scenario that can explain the puzzling HNC profiles is the one explored in simulations carried out by \cite{Chira-2014}. They performed the simulation of a cluster with cores embedded in filaments and make radiative transfer calculations of HCO$^+$ and HCN in several transitions (not HNC as in our observations). In their simulations, the gas in dense filaments in front of and moving toward a core, red-shifted from the observer line of sight, can be  sufficiently dense  and emit at low transitions to produce non-blue asymmetries in the observed line profiles (by making the red-shifted peak brighter). According to \cite{Chira-2014}, in irregular-collapsing cores infall signatures may not be evident in all molecular tracers at a given transition. In our case, HCO$^+$ shows infall signatures while HNC does not. In order to  confirm \cite{Chira-2014} predictions in collapsing cores, specific simulations for HNC and observations at higher transitions are necessary.

\section{Discussion}

\subsection{Prestellar Core Dynamics}

\subsubsection{Virial Equilibrium}

To calculate the virial mass and virial parameter, we used the emission from the optically thin, cold gas tracers N$_2$D$^+$ (3-2) and DCO$^+$ (3-2).
We fitted a Gaussian profile to each spectrum across the ALMA1 core. The average values of the velocity dispersion are 0.27 $\pm$ 0.03 km s$^{-1}$ for N$_2$D$^+$ and 0.27 $\pm$ 0.04 km s$^{-1}$ for DCO$^+$. Using the
observed velocity dispersion ($\sigma_v$) and the thermal velocity dispersion ($\sigma_{th}$) of the deuterated molecules, we can calculate the non-thermal component and assess how turbulent is the gas in the core. The non-thermal velocity dispersion ($\sigma_{nt}$) is calculated from $\sigma_v^2=\sigma_{th}^2+\sigma_{nt}^2$. The thermal velocity dispersion is given by $\sigma_{th} = (k_{\rm B}T/\mu m_H)^{1/2}$, with $\mu$ the molecular weight. Using the dust temperature of the clump (14 K), for both deuterated molecules $\sigma_{th}$ = 0.062 km s$^{-1}$ (same molecular weight, $\mu$). The Mach number ($\sigma_{nt}$/$\sigma_{th-H_2}$) is then 1.19 ($\sigma_{th-H_2}$ = 0.22 km s$^{-1}$), indicating that the gas in the core is transonic. 


The virial mass of ALMA1 is calculated using
\begin{equation}
\label{eq-virial}
M_{vir}=3\left(\frac{5-2n}{3-n}\right)\frac{\sigma_v^2R}{G},
\end{equation}
\noindent where $R$ is the effective radius of the core, $G$ is the gravitational constant, and $n$ is a constant whose value depends on the density profile of the core as function of the radial distance, $\rho(r) \propto r^{-n}$ \citep{MacLaren-1988}. Here we adopt a value of $n=1.8$, which has been found to be representative for high-mass star-forming regions \citep{Mueller-2002,Garay-2007}. The virial mass of ALMA1 is derived to be $M_{vir}=2.4\pm0.6$ M$_\odot$, and its virial parameter is $\alpha_{vir}=0.14\pm0.08$, assuming a 50\% uncertainty in the mass value. The low value of the virial parameter for ALMA1 suggests that turbulence is insufficient to maintain the stability of ALMA1 \citep{Pillai-2011,Tan-2013,Lu-2015,Zhang-2015,Ohashi-2016,Sanhueza-2017}. 

It is possible that magnetic fields play a role in the stability of this core \citep{Zhang-2014,Frau-2014}. The virial mass taking into account the contribution of a magnetic field support would be given by:
\begin{equation}
M_{B,vir}=3\frac{R}{G}\left(\frac{5-2n}{3-n}\right)\left(\sigma_v^2+\frac{1}{6}\sigma_A^2\right),
\end{equation}
\noindent where $\sigma_A$ is the Alfven velocity, and $n=1.8$ as in Equation \ref{eq-virial}. The Alfven velocity is given by $\sigma_A= B/\sqrt{4\pi\rho}$, where B is the magnitude of the magnetic field and $\rho$ is the core mass density \citep{Sanhueza-2017}. 

In order to reach virial equilibrium ($M_{B,vir}/M_{core}$ = 1), the magnetic field strength in ALMA1 would need to be of the order of 1.2$\pm0.4$ mG (assuming a 50\% uncertainty in the core's mass). Observations of the magnetic field have not been done toward \myclump. Such field strengths have been reported in hot molecular cores \citep{Zhang-2014}, and recently observed toward prestellar core candidates in the high-mass regime \citep{Beuther-2018}.

\subsubsection{A Relatively Massive Starless Core Accreting From Its Environment}
We estimated the core accretion rate from the HCO$^+$ (3-2) molecular line emission. The HCO$^+$ profile is blue peaked, characteristic of infall,  toward most of the area defining ALMA1.

To estimate the infall velocity we used the ``Hill5'' model \citep{devries-2005}. ``Hill5'' is a simple radiative transfer model that can reproduce the observed spectral asymmetries which are expected to arise in a contracting core. In this model, the core has a peak excitation temperature $T_{peak}$ at the center and at the near and far edges of the core it has a excitation temperature $T_0$. Thus, the core is modeled as a two layer slab, where the excitation temperatures increase linearly up to a peak temperature at the boundary between the two regions ($T_{peak}$), and then decreases linearly back to the initial temperature ($T_0$). For this model the free parameters to fit are (1) the peak excitation temperature ($T_{peak}$), (2) the velocity dispersion of the molecular line ($\sigma$), (3) the optical depth of the line ($\tau$), (4) the velocity of the cloud with respect to the Local Standard of Rest ($V_{LSR}$), and (5) the infall velocity of the gas in the core ($V_{in}$). This model may underestimate the infall velocity in some cases. However, the reliability of the model improves when the infall velocity is higher than the velocity dispersion of the line and when the line profile has a separated red-shifted peak. Both conditions met by our observations.

\begin{figure}
\begin{center}
\includegraphics[angle=0,scale=0.45, trim={0.5cm 0 1.5cm 0}, clip]{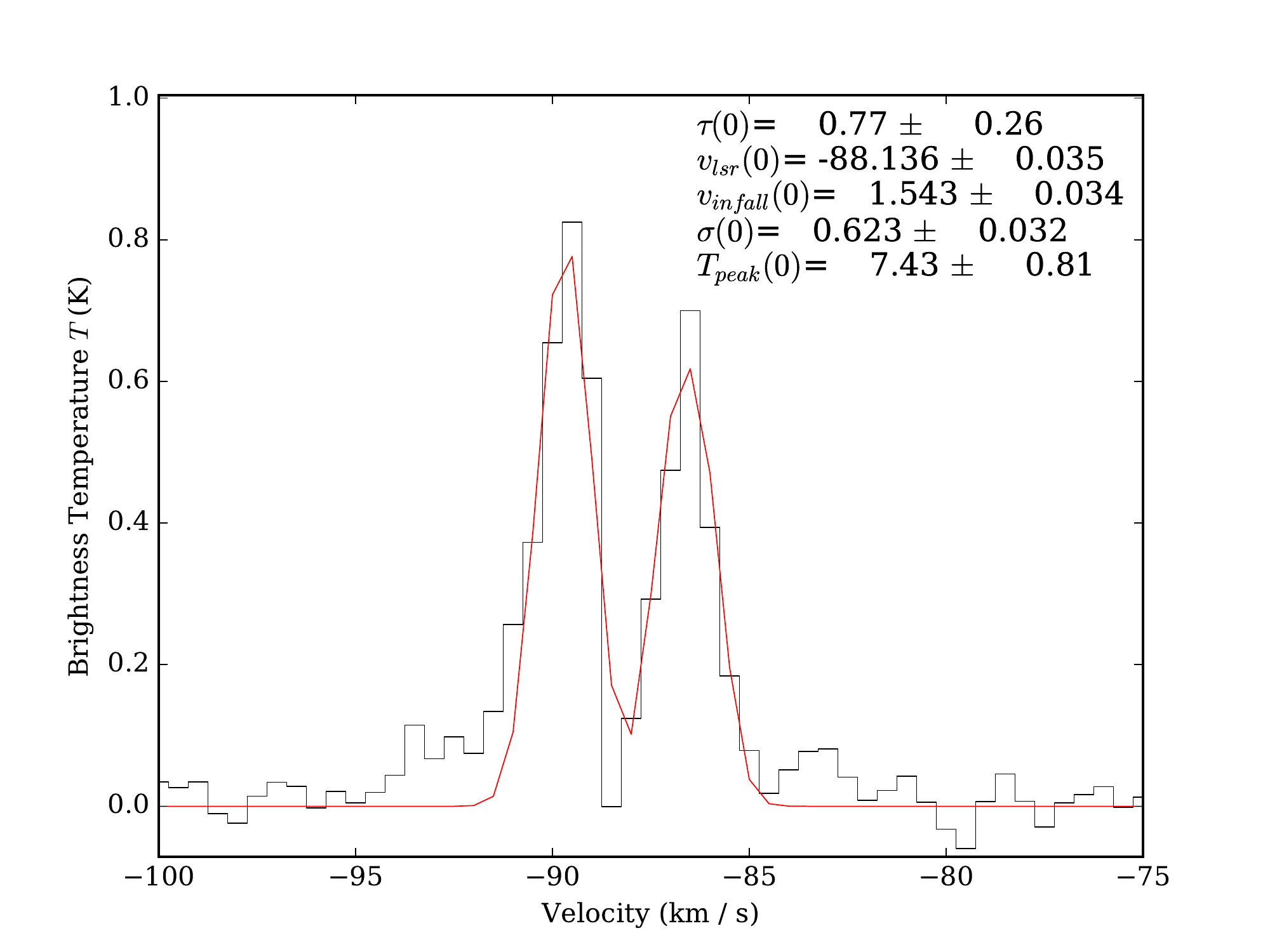}
\end{center}
\caption{Mean HCO$^+$ (3-2) spectrum across ALMA1 (black). In red is shown the Hill5 fit. In the upper right corner is shown the parameters fitted by the model.}
\label{hill5_spec}
\end{figure}

To determine the global accretion rate toward ALMA1, we fitted the average spectrum of the HCO$^+$ emission detected across ALMA1 using Hill5. To perform the fitting, we used the Hill5 model from the PySpecKit\footnote{http://pyspeckit.readthedocs.io} spectroscopic analysis toolkit \citep{Ginsburg-2011}. PySpecKit uses \texttt{lmfit} to perform the fit to the data to the Hill5 model. The uncertainties in the fitting are given by \texttt{lmfit} functions that explicitly explore the parameter space and determine confidence levels. As initial guesses of the fit we used $\tau$ ranging from 0.1 to 30, a $v_{LSR}$ between -89 and -86.6 km s$^{-1}$, $v_{in}$ between 0.1 and 4 km s$^{-1}$, $\sigma$ between 0.3 and 0.9 km s$^{-1}$, and $T_{peak}$ between 2 and 30 K. Figure \ref{hill5_spec} shows the spectra modeled by Hill5. Table 2 shows the values for ALMA1 of the $\tau$, $v_{LSR}$, $v_{in}$, $\sigma$, and $T_{peak}$ obtained by the model. In this table we also shows the parameters derived from the optically thin tracers detected toward this core. 

The mass infall accretion rate of the core was calculated via $\dot{{M}}=4\pi R^2\rho v_{in}$. For ALMA1 its mean density is $\rho=(8.5\pm2.7)\times10^5$ cm$^{-3}$, $v_{in}=1.54\pm0.03$ km s$^{-1}$ is the mean infall velocity across the core, and R=0.041 pc is its radius. Thus, we estimate a mass infall rate of (1.96$\pm 0.10) \times 10^{-3}$ M$_\odot$ yr$^{-1}$, which is an lower limit if we consider that the model can underestimate the infall velocity of the core. This accretion rate is comparable to values measured in high-mass star-forming regions \citep[e.g.,][]{Fuller-2005,Sanhueza-2010,Peretto-2013,Liu-2017, Liu-2013}

We compare this value to the simple case of assuming the core as a singular isothermal sphere. In this case the mass accretion rate of the core assuming turbulent motions and a star formation efficiency of 0.3 is given by
$\dot{M}\sim 0.441 \sigma^3/G$ \citep{mckee-tan-2003}, where $\sigma$ is the HCO$^+$ (3-2) velocity dispersion derived from the Hill5 fit. This simple model gives an accretion rate of $(2.5\pm0.4)\times 10^{-5}$ M$_\odot$ yr$^{-1}$, which is significantly low compared to the accretion rate obtained from the Hill5 model and to other regions of high-mass star formation \citep{mckee-tan-2003}.

\subsubsection{Prospects for High-Mass Star Formation in IRDC G331.372-00.116}

In spite of having all necessary physical properties to form high-mass stars, IRDC G331.372-00.116 has only one relatively massive core of 17.6 M$_\odot$. This core can be hardly considered  a high-mass prestellar core as the core accretion theory predicts \citep[see discussion in][]{Tan-2013,Tan-2014,Sanhueza-2017}. \cite{Tan-2017} suggests that in a similar IRDC, G028.23-00.19 \citep{Sanhueza-2013,Sanhueza-2017}, high-mass cores are absent because they have not yet formed. This could also be the case for G331.372-00.116 and high-mass prestellar cores could eventually form later. However,  ALMA1 is a relatively massive, sub-virialized core that has signs of accretion. This accretion can increase the mass of the core, making the scenario observed in IRDC G331.372-00.116 partially consistent with competitive accretion theories \citep{bonnell-2004,Wang-2010}, in which the core increases its mass over time to eventually gather sufficient mass to form a high-mass star.

However, the thermal Jeans mass ($M_J$) of the IRDC \myclump, given by
\begin{equation}
M_J = \frac{\pi^{5/2}}{6}\frac{\sigma_{\rm th}^3}{\sqrt{G^3\rho}}~,
\label{jeansmass}
\end{equation}
is only 2.8 \Msun\ (with $\sigma_{\rm th}$ the thermal velocity dispersion of the gas , dominated by H$_2$ and He, and $\rho$ the clump mass density). Thus, ALMA1 is 6 times more massive than the Jeans mass, which is high to agree with the competitive accretion theory unless the core has had sufficient time to grow. At the current accretion rate, a core with a Jeans mass would need 7.0$\times$10$^3$ yr to increase its mass to the current value of ALMA1. Our finding in ALMA1 is consistent with studies of more evolved massive IRDCs by \citet{Zhang-2009,Zhang-2011}, and \citet{Wang-2014} who report  cores of super Jeans masses.

The low turbulence in ALMA1, characterized by a transonic Mach number, results in a low virial parameter that suggests rapid collapse, similar to what is found in other prestellar cores embedded in high-mass 
star-forming clumps \citep{Sanhueza-2017,Lu-2018}. 
If we take into account magnetic field support, then ALMA1 may be maintained in virial equilibrium as the turbulent core accretion model requires \citep{mckee-tan-2003}. However, the rapid collapse is confirmed at high-angular resolution by the blue-red asymmetry characteristic of collapse. The accretion in ALMA1 is indeed more than 100 times faster than typically measured in low-mass star-forming cores and what is predicted by collapse of an isothermal sphere \citep{mckee-tan-2003,shu-1977}. 

Should magnetic field be negligible, the subvirial state of ALMA1 is supposed to cause collapse and fragmentation. Indeed, it is expected that cores that form high-mass stars will form binary system given that ~80\% of high-mass stars are found in binary systems \citep{Chini-2012}. For ALMA1, the thermal Jeans length corresponds to $\lambda_J=0.025$ pc. This size is one third of ALMA1 size and it is comparable to the synthesized beam of the observations (1\farcs2, 0.03 pc). Therefore, if ALMA1 collapses it might fragments only into a few cores of roughly the size of the thermal Jeans length. 

Whether or not ALMA1 fragments into smaller units, there is a high probability that the core will form at least one high-mass star. For ALMA1 the core free fall time is $t_{ff}=\sqrt{3\pi/(32G\rho)}=3.3\times10^4$ yr. If ALMA1 continue accreting material from its environment, which can be possible given the large reservoir of gas within \myclump, at its current accretion rate it can grow to reach a $\sim$82 M$_\odot$ core on a timescale of its free fall time (assuming a constant accretion rate over the whole time). Thus, it is highly likely that ALMA1 is the initial cocoon of what will be at least one high-mass star in the future.

We suggest that ALMA1 will likely form a high-mass star and that there may be no need to wait for the formation of a high-mass prestellar core in IRDC G331.372-00.116 in order to form a high-mass star. However, there remain two unanswered questions that we are going to answer in our future studies. First, would ALMA1 indeed fragment at high angular resolution observations? Second, is ALMA1 globally fed by/through filamentary accretion?. We also suggest that cores that will ultimately form high-mass stars begin in a sub-virial state with a low level of turbulence.  

\section{Conclusions}
We studied the internal structure of the 70 $\mu$m dark IRDC \myclump~with ALMA. Although \myclump~satisfy all the conditions to form a high-mass star, the most massive core, ALMA1, has only a mass of 17.6 M$_\odot$. 

The virial mass of ALMA1 is 2.4$\pm0.6$ M$_\odot$ and its virial parameter is $0.14\pm0.08$, suggesting that the core is subvirial and turbulence alone cannot halt collapse. Table 3 summarizes all the properties derived for ALMA1.  

We found evidence of infall toward ALMA1, as shown by the blue-peaked profiles exhibited by HCO$^+$ molecular line emission. Using the Hill5 model we estimated an infall velocity of 1.54$\pm0.03$ km s$^{-1}$ and a mass infall accretion rate of (1.96$\pm0.10)\times10^{-3}$ M$_\odot$ yr$^{-1}$. 
If ALMA1 continue to gather material from its environment at the current rate, this core can reach to $\sim82$ M$_\odot$ on time scales comparable to the core free-fall time. 

Overall, our observations suggest that ALMA1 will likely form a high-mass star, and the starting point for such high-mass star is a sub-virial core of an intermediate mass with low turbulence that will increase it mass via accretion.

\begin{center}
\begin{deluxetable*}{lcccccccccc}
\tabletypesize{\footnotesize}
\tablecaption{Properties of the sub-structures detected towards \myclump \label{tbl-clumps}}
\tablewidth{0pt}
\tablehead{
\colhead{Core} & RA & DEC &Radius & Peak Flux & Integrated Flux & N(H$_2$) & Mass & n(H$_2$) & $\Sigma$ \\
\colhead{$\mathrm{Name}$} & (J2000) & (J2000) & $\mathrm{(pc)}$ & $\mathrm{(mJy\,beam^{-1})}$ & $\mathrm{(mJy)}$ & $\mathrm{(\times10^{23}~cm^{-2})}$ & $\mathrm{(M_{\odot})}$ & $\mathrm{(\times10^6~cm^{-3})}$ & $\mathrm{(g\,cm^{-2})}$ \\
}
\startdata
ALMA1 & 16:11:33.30 & -51:35:23.3	 & 0.041 & 4.07 & 25.9 & 2.95 & 17.6 & 0.85 & 0.65 \\
ALMA2 & 16:11:32.68 & -51:35:21.2	 & 0.019 & 1.62 & 3.27 & 1.18 & 2.22 & 1.19 & 0.41 \\
ALMA3 & 16:11:34.64 & -51:35:17.1	 & 0.013 & 1.13 & 1.30 & 0.81 & 0.88 & 1.38 & 0.33 \\
ALMA4 & 16:11:33.14 & -51:35:17.1	 & 0.014 & 1.65 & 2.36 & 1.20 & 1.60 & 1.83 & 0.49 \\
ALMA5 & 16:11:33.31 & -51:35:15.1	 & 0.019 & 1.46 & 2.83 & 1.06 & 1.92 & 0.97 & 0.34 \\
ALMA6 & 16:11:33.93 & -51:35:10.2	 & 0.023 & 1.51 & 4.81 & 1.09 & 3.26 & 0.88 & 0.38 \\
ALMA7 & 16:11:33.43 & -51:35:10.5	 & 0.018 & 1.64 & 3.39 & 1.19 & 2.30 & 1.37 & 0.45 \\
ALMA8 & 16:11:33.14 & -51:35:09.7	 & 0.012 & 1.32 & 1.30 & 0.95 & 0.88 & 1.70 & 0.38 \\
ALMA9 & 16:11:34.04 & -51:35:04.1	 & 0.030 & 1.65 & 6.84 & 1.19 & 4.64 & 0.59 & 0.33 \\
ALMA10& 16:11:34.05 & -51:34:59.5	 & 0.034 & 2.59 & 10.9 & 1.88 & 7.40 & 0.66 & 0.41 \\
ALMA11& 16:11:34.46 & -51:34:56.4	 & 0.014 & 1.31 & 1.66 & 0.95 & 1.13 & 1.43 & 0.37 \\
ALMA12& 16:11:33.85 & -51:34:54.0	 & 0.017 & 1.29 & 1.97 & 0.93 & 1.34 & 0.87 & 0.28 \\
ALMA13& 16:11:34.36 & -51:34:53.7	 & 0.012 & 1.13 & 1.20 & 0.82 & 0.81 & 1.56 & 0.35 \\
ALMA14& 16:11:34.84 & -51:34:50.7	 & 0.019 & 1.25 & 2.52 & 0.91 & 1.71 & 0.90 & 0.31 \\
ALMA15& 16:11:34.29 & -51:34:51.1	 & 0.014 & 1.26 & 1.71 & 0.91 & 1.16 & 1.32 & 0.35 \\
ALMA16& 16:11:34.05 & -51:34:51.0	 & 0.017 & 1.45 & 2.42 & 1.05 & 1.64 & 1.24 & 0.38 \\
ALMA17& 16:11:33.28 & -51:34:50.5	 & 0.015 & 3.01 & 3.82 & 2.18 & 2.60 & 2.55 & 0.71 \\
ALMA18& 16:11:33.64 & -51:34:49.5	 & 0.017 & 2.09 & 3.63 & 1.52 & 2.47 & 1.80 & 0.56 \\
ALMA19& 16:11:34.88 & -51:34:48.7	 & 0.013 & 1.65 & 1.75 & 1.20 & 1.19 & 1.98 & 0.46 \\
ALMA20& 16:11:34.02 & -51:34:46.7	 & 0.041 & 2.06 & 15.7 & 1.49 & 10.7 & 0.54 & 0.41 \\
ALMA21& 16:11:33.43 & -51:34:36.3	 & 0.019 & 1.23 & 2.78 & 0.89 & 1.88 & 0.94 & 0.33 \\
ALMA22& 16:11:32.61 & -51:34:34.7	 & 0.041 & 1.60 & 13.9 & 1.16 & 9.42 & 0.46 & 0.35 \\
\enddata
\end{deluxetable*}
\end{center}

\begin{deluxetable*}{lccccccccccccccc}
\tabletypesize{\footnotesize}
\tablecaption{Summary of the line parameters of the optically thin tracers observed towards ALMA1, and the derived parameters from the Hill5 model.}
\tablewidth{0pt}
\tablehead{
\colhead{} && $\tau$  & $v_{in}$ & T$_{peak}$& $\sigma$  
 & $v_{LSR}$& \\
\colhead{$\mathrm{}$} && $\mathrm{}$ & $\mathrm{(km\,s^{-1})}$ & (K)  & $\mathrm{(km\,s^{-1})}$ & $\mathrm{(km\,s^{-1})}$ & Method
}
\startdata
HCO$^+$   &J=3-2& 0.77$\pm$0.26 & 1.54$\pm$0.03 & 7.43$\pm$0.81& 0.62$\pm$0.03 & -88.14$\pm0.04$ &Hill5\\	
N$_2$D$^+$ &J=3-2& & & 0.91$\pm$0.09	& 0.27$\pm$0.03	& -87.69$\pm$0.03&Gaussian Fit\\
DCO$^+$&J=3-2&&&	0.66$\pm$0.08	& 0.27$\pm$0.04	& -87.71$\pm$0.04&Gaussian Fit\\
C$^{18}$O&J=2-1&&&	0.67$\pm$0.05	& 0.68$\pm$0.06	& -87.37$\pm$0.06&Gaussian Fit \\
\enddata
\end{deluxetable*}

\begin{deluxetable}{ll}
\tabletypesize{\footnotesize}
\tablecaption{Summary of physical parameter derived for ALMA1}
\tablewidth{0pt}
\tablehead{
\colhead{Parameter} & Value 
}
\startdata
Flux (267 GHz) & 26 mJy\\
$r_{eff}$ & 0.041$\pm$0.004 pc\\
$M$ & 17.6$\pm$8.8 M$_\odot$\\
$M_{vir}$ & 2.4$\pm$0.6 M$_\odot$\\
$\alpha_{vir}$ & 0.14$\pm$0.08\\
n(H$_2$) & (8.5$\pm$2.7)$\times 10^5$ cm$^{-3}$\\
$\Sigma$ & 0.65 gr cm$^{-2}$\\
$\mathcal{M}$& 1.19\\
$\dot{M}$ & (1.96$\pm$0.10)$\times10^{-3}$ M$_\odot$ yr$^{-1}$\\
t$_{ff}$ & 3.3$\times 10^4$ yr\\
\enddata
\end{deluxetable}

  \acknowledgements
We thank the anonymous referee for their comments that have greatly improved the quality of this paper. Y.C., A.E.G., and P.S. gratefully acknowledge the support 
from the NAOJ Visiting Fellow Program to visit the National Astronomical Observatory of Japan in November-December 2016.  Y. C. acknowledges assistance from Allegro, the European ALMA Regional Center node in the Netherlands. A.E.G. thanks FONDECYT N$^{\rm o}$ 3150570. G. G. acknowledges support from Conicyt project PFB-06. Data analysis was in part carried out on the open use data analysis computer system at the Astronomy Data Center, ADC, of the National Astronomical Observatory of Japan. This paper makes use of the following ALMA datasets: ADS/JAO.ALMA\#2013.1.00234.S and ADS/JAO.ALMA\#2015.1.01539.S. ALMA is a partnership of ESO (representing its member states), NSF (USA) and NINS (Japan), together with NRC (Canada) and NSC and ASIAA (Taiwan) and KASI (Republic of Korea), in co- operation with the Republic of Chile. The Joint ALMA Observatory is operated by ESO,AUI/NRAO and NAOJ. This research made use of astrodendro, a Python package to compute dendrograms of Astronomical data (http://www.dendrograms.org/)
 
\vspace{5mm}
\facilities{ALMA}


\software{CASA 4.7.2}

\bibliography{bibliografia}

\appendix
\section{Automatic cleaning algorithm}
\label{appendixyclean}

We have developed an automatic cleaning algorithm for imaging data cubes, \texttt{yclean}. The algorithm is written in \texttt{python} and make use of the \texttt{tclean} task in the CASA software. \texttt{yclean} creates individual masks for every channel of a line data set  through an iterative process in order to obtain the optimal mask per channel for cleaning.

The \texttt{yclean} algorithm determines a threshold to mask every channel that depends on the value of the secondary lobe of the PSF, the rms of the image, and the maximum value of the image residual. First, it creates a dirty beam image with the \texttt{tclean} task in CASA. From the dirty .psf image, the algorithm determines the value of the secondary lobe (\texttt{secondaryLobe}) of the PSF map. From the .im image, it calculates the rms of the map (\texttt{rmsMap}) by computing the average rms value (using CASA \texttt{imstat} task) of a subset of channels located in a line free region of the map (usually near the edges of the spectral window). Finally, from the .residual image we obtain the maximum value (\texttt{maxResidual}). With the values of \texttt{secondaryLobe}, \texttt{rmsMap}, and \texttt{maxResidual}, a threshold is defined to create a mask that will be used with \texttt{tclean} (\texttt{limitLevelSNR}). 

The \texttt{limitLevelSNR} is given by:
\begin{equation}
\texttt{limitLevelSNR}=\frac{\texttt{maxResidual}}{\texttt{rmsMap} \times\texttt{secondaryLobe}}
\end{equation}

After computing the first mask, a loop begins that finishes when the measured value of \texttt{limitLevelSNR} is greater than 2. In each loop, the following actions are performed:
\begin{itemize}
\item Create a mask. 
\item Combine the new mask with the mask used in the previous iteration.
\item Use \texttt{tclean} to clean the image using the mask previously created. In this step, the image is cleaned until it reaches a threshold given by \texttt{limitLevelSNR}$\times$\texttt{rmsMap}.
\item Calculate a new value of \texttt{limitLevelSNR}.
\end{itemize}

Once the loop ends, the \texttt{yclean} script performs a final clean using \texttt{tclean} with a cleaning stopping threshold given by 2$\times$\texttt{rmsMap} and using a final mask that combines the previous mask with a new mask created with a threshold of 4$\times$\texttt{rmsMap}. Finally, after the cleaning is done, all the products are exported into fits format.

More documentation about the \texttt{yclean} script, along with the necessary modules to run it, are available at www.yanettcontreras.com/yclean.html. A frozen version of the script can also be found at zenodo (10.5281/zenodo.1216881).

\end{document}